# A REAL-TIME ENERGY MONITOR SYSTEM FOR THE IPNS LINAC*

J.C. Dooling, F. R. Brumwell, M.K. Lien, G. E. McMichael, ANL, Argonne, IL 60439, USA

*Abstract*

Injected beam energy and energy spread are critical parameters affecting the performance of our rapid cycling synchrotron (RCS). A real-time energy monitoring system is being installed to examine the H$^-$ beam out of the Intense Pulsed Neutron Source (IPNS) 50 MeV linac. The 200 MHz Alvarez linac serves as the injector for the 450 MeV IPNS RCS. The linac provides an 80 μs macropulse of approximately $3 \times 10^{12}$ H$^-$ ions 30 times per second for coasting-beam injection into the RCS. The RCS delivers protons to a heavy-metal spallation neutron target for material science studies. Using a number of strip-line beam position monitors (BPMs) distributed along the 50 MeV transport line from the linac to the RCS, fast signals from the strip lines are digitized and transferred to a computer which performs an FFT. Corrections for cable attenuation and oscilloscope bandwidth are made in the frequency domain. Rectangular pulse train phasing (RPTP) is imposed on the spectra prior to obtaining the inverse transform (IFFT). After the IFFT, the reconstructed time-domain signal is analyzed for pulse width as it progresses along the transport line. Time-of-flight measurements of the BPM signals provide beam energy. Finally, using the 3-size measurement technique, the longitudinal emittance and energy spread of the beam are determined.

## 1 INTRODUCTION AND MOTIVATION

The Intense Pulsed Neutron Source (IPNS) accelerator system is equipped with a number of strip-line, beam position monitors (BPMs) along the 40-m transport line from the 50 MeV Linac to the Rapid Cycling Synchrotron (RCS). Operating at 30 Hz, the RCS delivers 450 MeV protons to a heavy metal target generating spallation neutrons for material science research. Here we describe how signals from the first four (upstream) BPMs in the 50 MeV line are used to determine bunch width, energy, and energy spread in the beam. Injected beam energy spread plays an important role in determining the stability of circulating charge within a synchrotron. Advancements in the speed of sampling oscilloscopes and the rapid increase in processing power available from personal computers allow for real-time measurement of the output microbunch shape from the linac. As the bunch travels along the transport line, its longitudinal size grows due to energy spread within the bunch. The growth in bunch length can be monitored with the stripline BPMs and the energy spread determined.[1]

The IPNS Alvarez, drift tube linac (DTL) began operation in 1961 as the injector for the Zero Gradient Synchrotron (ZGS). In 1981 after the ZGS program ended, the linac became the injector for IPNS RCS. The linac typically delivers $3.5\text{-}3.7 \times 10^{12}$ H$^-$ ions to the RCS during an 80-μs macropulse. During the early days of the linac, the energy spread was measured to be 0.37 MeV. From numerical modeling, the capture efficiency of the RCS at injection is optimized near a momentum spread of 0.3 percent or approximately 0.3 MeV. Instabilities arise if the energy spread is too low, whereas high losses occur if the spread is too high; in either case, RCS efficiency is reduced. A shift in energy during the macropulse effectively acts to increase energy spread during injection into the RCS.

## 2 EXPERIMENTAL ARRANGEMENT

The Energy Spread and Energy Monitor (ESEM) diagnostic is presented schematically in Figure 1. The upstream electrode of the first BPM (BPM 1) is located 5.455 m from the output flange of the last DTL tank. The set of four BPMs included in the ESEM cover a distance of 16.627 m along the beam path. This distance is sufficient to allow observable growth in the longitudinal size of the bunch without interference from the return signal generated at the downstream electrode of the BPM.

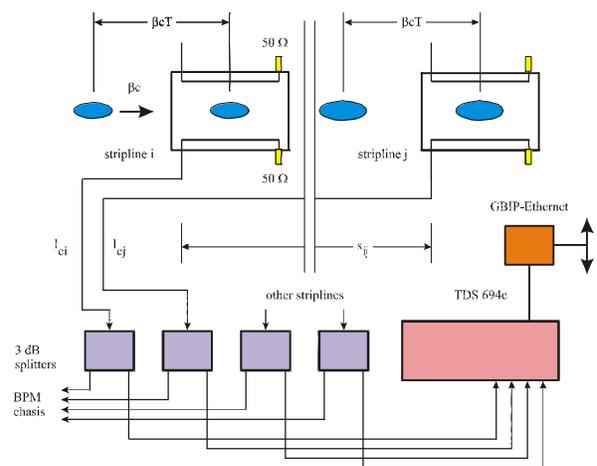

Figure 1. The Energy Spread and Energy Monitor.

*This work is supported by the US DOE under contract no. W-31-109-ENG-38.

# 3 DESCRIPTION OF MEASUREMENTS

## 3.1 Pulsewidth

Time signals from the BPMs are recorded on a Tektronix model TDS694c oscilloscope. The TDS694c has a 3 GHz bandwidth and samples 4 channels independently at $10^{10}$ samples/sec (10 GS/s). The oscilloscope is controlled and data transferred via a GPIB-to-Ethernet interface. This allows an office PC on the network to communicate with the oscilloscope. After receiving the initial trigger pulse from the chopper, the oscilloscope waits for a controlled amount of time before triggering its four channels. The delay time is adjustable to allow for temporal examination of the macropulse. Once the delay period has expired, oscilloscope triggering is enabled for the next zero crossing detected on channel one. When the zero crossing is detected, all four channels are triggered simultaneously. The trigger time establishes the reference time for energy measurements. A sampled waveform from BPM 1 is presented in Figure 2.

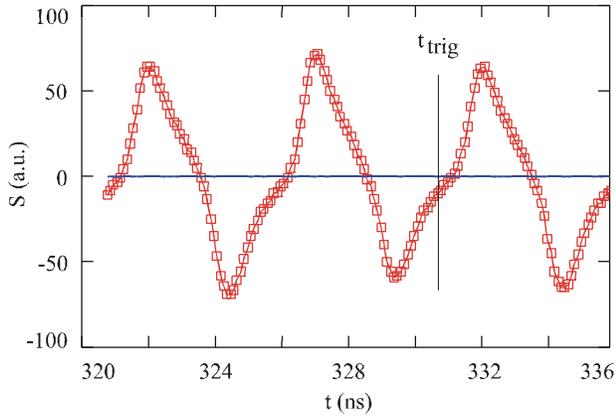

Figure 2. Sampled waveform from BPM1.

Once signals from the BPMs have been collected and transferred to the PC, the data can be displayed and analyzed. Data analysis is implemented using Visual Basic. Analysis of the signals begins with rebinning the data to obtain data sets with $2^N$ samples as required for the FFT[2].

In order to examine the same bunch on all four BPMs and account for cable delays, a data window of 100 ns or longer is required. It is also desirable have good frequency response for parameters derived from the entire FFT. A 100 ns window allows us to look at parameters derived from the FFT up to 5 MHz (Nyquist). A period of 100 ns is a good compromise between resolution in the FFT and frequency response of the derived parameters. Extending the sample window requires more time but also provides at cleaner signal.

After carrying out the FFT, the spectra are converted to dBm and corrected for cable attenuation and insertion losses. RG-213 coax cable is employed to carry the beam signals from the BPM to the oscilloscope. An attenuation correction[3] that is a function of frequency and cable length is applied to each spectrum. Finally, when performing the inverse transform, Fourier filtering is performed to remove low-level signals lost in the noise. The pulsewidth is determined by evaluating the following time function,

$$f(t) = \frac{1}{\sqrt{N_b}} \sum_{n=1}^{N_{max}} F_n \cos(n\omega t + \phi_n)$$

where $F_n$ is the amplitude of the $n^{th}$ harmonic of the beam bunch determined from the FFT. Employing rectangular pulse train phasing (RPTP), $\phi_n$ may be expressed as,

$$\phi_n = n\omega_0 (t_0 + \frac{t_b}{2})$$

The original phase is also available from the FFT by taking the inverse tangent of the ratio of imaginary to real parts of the signal. Figure 3 presents the cable-corrected FFT spectrum for the time data given in Figure 2. Also indicated in the figure is the threshold level. Finally, the reconstructed time signals using both RPTP and original phasing are given in Figure 4. The pulse shape in either case is similar to the "bipolar doublet" described by Shafer[4] and shown by Kramer[1].

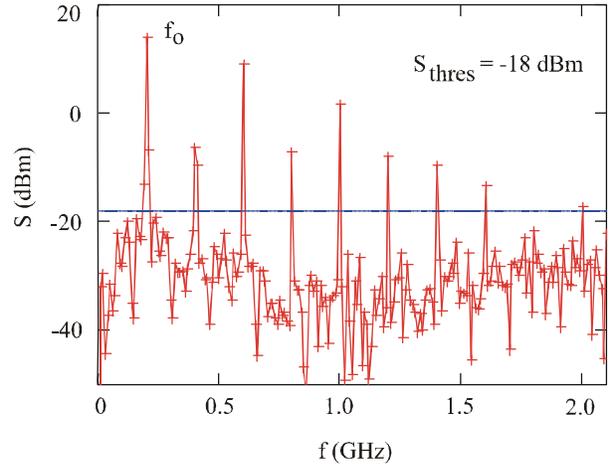

Figure 3. Corrected FFT spectrum for data given in Fig. 2.

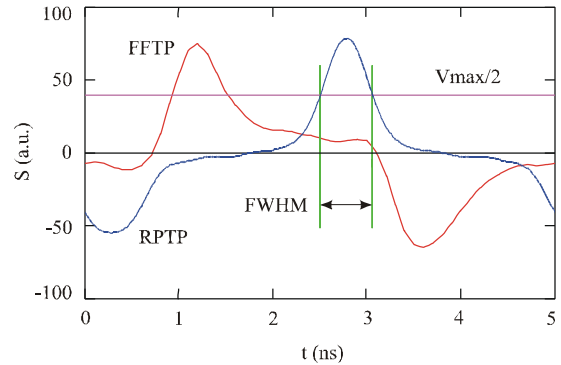

Figure 4. The reconstructed time signals.

## 3.2 Energy

Energy is determined by tracking a single microbunch through the four BPMs. Time-of-flight (TOF) measurements require accurate knowledge of the cable lengths between the BPMs and the oscilloscope, and the distance travelled by the beam along the transport line. Cable lengths are determined by TDR. The path length of the beam is measured directly or taken from survey data.

The energy is calculated from the measured velocity of the bunch. Velocity is determined by measuring the TOF and correcting for signal propagation delays along the cables. The time required for the signals from a given bunch to reach the oscilloscope may be expressed as,

$$t_j = \frac{l_{cj}}{v_p} + \frac{s_{1j}}{bc}.$$

where j=2, 3, and 4 and $t_1 = l_{c1}/v_p$. Taking $t_1$ as the reference time, the bunch velocity in terms of the time difference between arrival on channel 1 and channel j is given as,

$$b_{1j}c = \frac{v_p s_{1j}}{t_{j1} v_p - l_{cj} + l_{c1}}$$

where, $v_p$ is the phase velocity in the cable or transmission line, $l_{cj}$ is the length of cable j, $s_{1j}$ is the distance the bunch travels from BPM 1 to BPM j, and $t_{j1}=t_j-t_1$. The sensitivity of the measurement increases with distance between channel 1 and channel j. When calculating the energy, the analysis detects the leading edge of the bunch at the half-height of the pulse. To obtain a measure of energy from bunch center-to-center, the pulsewidth of the bunch at each location must be accounted for.

## 3.3 Energy Spread

Energy spread is obtained by the three-size measurement technique[5] neglecting space charge. The advantage of this method is that it provides the longitudinal Twiss parameters and therefore gives an indication of axial focussing (i.e., rf fields) at the output of the DTL. The section of beam line between BPMs 2 and 3 includes two π/6 sector dipole magnets, both with 0.8 m radius of bend. Though the dipoles cannot affect the energy or energy spread of the beam, they do cause a change in the longitudinal size of the beam through dispersion and coupling with the horizontal plane. This effect is introduced into the longitudinal transfer matrix by adjusting the path length between BPM 2 and 3. The dipole length factor, $f_d$, can be included in the drift-space transfer matrix $R_d(s)$. Labeling the three drift spaces between the four BPMs as $s_1$, $s_2$, and $s_3$, the transfer matrix between BPM 1 and 3 for the 123 measurement can be written as,

$$R_3(s_1, s_2) = R_{2d}(s_2 - 2f_d l(r))R_{d1}(s_1)$$

For the 124, 134, and 234 measurements, the respective expressions for $R_3$ become,

$$R_3(s_1, s_2, s_3) = R_{d2}(s_2 + s_3 - 2f_d l(r))R_{d1}(s_1)$$
$$R_3(s_1, s_2, s_3) = R_{d2}(s_3)R_{d1}(s_1 + s_2 - 2f_d l(r))$$
$$R_3(s_2, s_3) = R_{d2}(s_3)R_{d1}(s_2 - 2f_d l(r))$$

In theory, all four permutations should generate the same result; in reality, $f_d$ may be varied to provide the least rms error. Initial data indicate $f_d$ is not a fixed parameter but varies significantly from sample to sample within a range of ±4. Negative values generally correlate with lower energy spread, whereas the highest positive values tend to correspond to the largest energy spread.

## 4 RESULTS AND DISCUSSION

The maximum average current that the RCS has achieved is 15.5 µA, reached in May of this year prior to the annual summer shutdown. Since restarting the machine on August 1st, maximum current has been limited to about 14.8 µA. The ESEM diagnostic was first put into service in May and observed some of the higher current operations. ESEM energy data recorded in May and August are presented in Figure 5.

ESEM energy data suggest that a greater fluctuation in linac energy during the August 2000 run may be limiting the efficiency of the RCS. Further tests are being planned to determine the validity of this supposition.

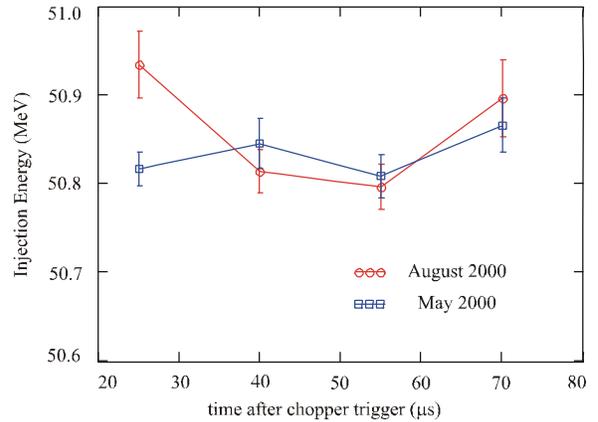

Figure 5. ESEM linac energy data.